# Magnetic Pillar Induced Poiseuille-like Flow in Microfluidic Channels with Viscous and Viscoelastic Fluids

Charles Paul Moore[1,2*], Stefan Rouach[2], Marine Le Goas[2], Sandra Lerouge[2], Nicolas Tsapis[3], Jérôme Fresnais[1] and Jean-François Berret[2*]

[1]*Sorbonne Université, CNRS, Laboratoire de Physicochimie des Electrolytes et Nanosystèmes Interfaciaux, PHENIX, UMR8234, F-75252 Paris Cedex 05, France*
[2]*Université Paris Cité, CNRS, Matière et systèmes complexes, 75013 Paris, France*
[3]*Université Paris-Saclay, CNRS, Institut Galien Paris-Saclay, 91400 Orsay, France*

**Abstract**
Mucociliary clearance in mammals serves as the primary defense mechanism for removing particulate matter deposited in the pulmonary airways. Dysfunctions in this process are linked to serious respiratory diseases and can hinder effective drug delivery to the lungs. Microfluidic systems have emerged as a promising alternative for replicating lung functions in non-cellular physiological environments, offering a simpler and more controllable approach compared to *in vivo* and *in vitro* assays. Here we present a microfluidic platform featuring a closed-loop circular microchannel, integrating thousand 75 µm-high magnetic pillars arranged in a square array. Made of polydimethylsiloxane and loaded with iron microparticles, the pillars are studied using scanning electron microscopy and magnetometry; their internal structure and bending response to a magnetic field are quantitatively analyzed. Using a combination of experimental data and finite element simulations, we found that the magnetic torque induced by permanent magnets dominates over magnetic force, generating fluid flow in the microchannel. Under the application of a rotating field, the time-dependent deflection of the pillars closely mimics the behavior of lung cilia, exhibiting alternating recovery phases and rapid whip-like movements. The velocity profiles of viscous and viscoelastic fluids are examined, and shown to display Poiseuille-type flow. By varying the viscosity of the fluids across four orders of magnitude, we identified a transition in propulsion regimes between viscous and elastic-driven flows. This active microfluidic platform offers a promising approach for modeling mucociliary clearance in drug delivery applications.

# 1 – Introduction

Humans inhale approximately 10m³ of air daily, carrying particulate matter, much of which deposits in the pulmonary airways. To counteract this, the respiratory system relies on mucociliary clearance (MCC) as a key defense mechanism in the tracheobronchial region to mitigate the effects of inhaled particles and pathogens. The MCC integrates a 5–10 µm thick mucus layer with the coordinated beating of cilia, which propel mucus from the respiratory tract and direct it into the digestive system for clearance.[1-3] Cilia are versatile structures found across many living systems, aiding for lubrication or enabling motility in organisms like protozoa and some aquatic invertebrates.[4-8] In lung diseases such as ciliary dyskinesia and cystic fibrosis however, disruptions in cilia coordination or changes in mucus rheology can impair respiratory function. These issues, compounded by alterations in mucus composition, can hinder effective drug delivery to the airways.[2,9-12] In the context of local drug delivery to the lungs, advancing therapeutic strategies remains a critical challenge in medical engineering and personalized medicine.[13,14]

To overcome the limitations of *in vivo* studies in the formulation of lung-specific drugs, *in vitro* experiments on reconstituted human bronchial epithelium were conducted, allowing for *in situ* study of changes in mucus rheology and the observation of cilia efficiency and coordination.[15-18] These approaches also enable the study of mucus flow and the evaluation of drug diffusion into the mucus layer.[12] However, in air-liquid interface culture, cilia can sometimes produce non-physiological flows, complicating their interpretation in the context of bronchial function. Furthermore, in these two-dimensional models, synchronized ciliary coordination is observed but is limited to patches approximately 20 µm in diameter. Efficient long-range unidirectional mucus flow, critical for clearance has been noted when both ciliary density and mucus viscosity are sufficiently high.[16] Despite these issues, reconstructed human bronchial epithelium in 2D remains a valuable model for studying the interface between air, mucus, and lung cells.

To avoid the constraints inherent to cell manipulation, synthetic ciliated substrate alternatives have been developed in parallel using established microfluidic and soft lithography methods.[19-23] Pulmonary cilia are 7 µm long and 200 nm in diameter, with a surface density of 200-300 cilia per cell.[24] These cilia exhibit a progressive, coordinated and asymmetric movement, called a metachronal wave, and beat at a frequency of 10-20 Hz, leading in humans to mucus clearance at a speed of 40-90 µm s$^{-1}$.[3,25] To date, research in this field has focused not on replicating ciliated substrates with exact physiological characteristics of density, aspect ratio, or size, but on developing responsive substrates that can mimic MCC functions.[26,27] To achieve this, researchers took inspiration from elastomeric pillar arrays, developed about 20 years ago, designed for measuring micrometer-scale forces or crafting superhydrophobic substrates.[28-30] These ciliated substrates, typically composed of PDMS or poly(acrylamide) were however not suitable for remote actuation.[31,32] For controlled deflection, active particles—often magnetic—are incorporated into the elastomer, enabling pillar orientation on demand.[19-21,23,27,33-40] Magnetic particles commonly incorporated into elastomers include iron (Fe$^0$) or iron oxide particles (maghemite $\gamma$-Fe$_2$O$_3$, magnetite Fe$_3$O$_4$), with sizes ranging from a few nanometers to several micrometers.[34] Notably, 0.5 µm iron carbonyl microparticles (hereafter noted FeMPs) offer an excellent balance of high magnetization, chemical stability, and cost-effectiveness.[41]

Among the various devices recently developed for microfluidic applications, a series of high-performance microchips has been designed to efficiently propel viscous liquids in confined environments. Over the past decade, significant advancements have been achieved, including the miniaturization of pillars,[27,38,40] increased flow rates,[20] programmable pillar movements,[37-40] and the generation of metachronal waves reminiscent of lung ciliated cells.[20-22,37,38] However, limited attention has been given to studying velocity profiles in closed microfluidic chips or exploring the rheological properties of the fluids in motion. In addition, the above-mentioned studies commonly rely on the local application of magnetic fields, generated by either electro- or permanent magnets, which produce fields between 10 and 100 mT at the pillar level. Yet, such magnetic fields are inherently non-homogeneous, with their amplitude and orientation varying spatially. This spatial variation induces both magnetic forces and torques on the system, which can influence chip performance. Despite their critical role in optimizing pillar density and minimizing dipolar interactions, the relative contributions of these effects have been inadequately examined.[26,41,42] The present study seeks to address these gaps by investigating key factors such as the nature of induced flows, the influence of fluid viscoelasticity, and the dynamics of magnet-pillar interactions *via* physics simulations.

Here, we fabricated a circular closed-loop microchannel using soft lithography, incorporating an array with approximately 1067 magnetic pillars arranged on a square grid and defining the actuation zone.[20] These pillars loaded with iron carbonyl particles are characterized in terms of composition and density, and their mechanical response to a magnetic field is analyzed quantitatively. Using finite element simulations, combined with experimental data, we model the impact of the magnetic field on pillar bending and find that the magnetic torque dominates over the magnetic force. The pillars are activated by the field created by rotating permanent magnets, producing fluid flow in the microchannel. Velocity fields studied for viscous and viscoelastic fluids over more than 4 orders of magnitudes show Poiseuille-type profiles in the channel, as well as a propulsion regime transition between viscous and elastic driven flows.

## 2 – Results and discussion
### 2.1 - Pillar Characterization
#### 2.1.1 - Pillar fabrication
Flexible cylindrical pillar arrays were fabricated using a soft lithography.[43] The fabrication process involves creating an initial mold of the pillar structure, which is converted into a counter-mold. This counter-mold is subsequently filled with polydimethylsiloxane (PDMS) polymer and cured. The resulting design features approximately 1 cm² substrates with pillars oriented perpendicular to the surface, having nominal diameters $d_{Pillar}$ and heights $l_{Pillar}$ of 20 µm and 75 µm respectively.[41,44] A scanning electron microscopy (SEM) image of an array is displayed in Fig. 1a. The inset in the figure provides a close-up view of a pillar, revealing a slight taper with a base diameter of 19.5 µm and a tip diameter of 22.1 µm, and a length of 74.4 µm, in good agreement with the nominal values. A small bulge, resulting from mold construction, is visible 25 µm from the base. To enable remote actuation, the pillars were made magnetically responsive, allowing them to bend in response to a magnetic field. This was achieved by incorporating iron carbonyl microparticles (FeMPs) into the PDMS matrix during fabrication. The FeMPs, as characterized by TEM, displayed a size distribution centered around a median value $d_{FePM}$ of 0.69 µm with a dispersity of 0.46 (Supporting Information

S1). The dispersity is defined as the ratio between the standard deviation and the average diameter. The FeMPs were suspended in ethanol and deposited onto a counter-mold micro-well array.[45] After evaporating the excess ethanol, the wells were filled with PDMS, which, once cured, formed a hybrid polymer composite. The fabrication method is summarized in Fig. 1b.

### 2.1.2 - Pillar iron content

SEM combined with energy-dispersive X-ray (EDX) analysis was conducted to confirm the presence of FeMPs within the array. In Fig. 1c, dispersed white dots are visible in the upper regions of the pillars, indicating the presence of FeMPs embedded in the PDMS matrix. Statistical analysis of FeMP distribution within the pillars ($n$ = 48) shows that particles are primarily concentrated in the upper part, at a height $h_{FeMP}$ where $h_{FeMP}/l_{Pillar}$ = 0.5 ± 0.1 (SD). The two insets on the right display the SEM image of a pillar lying on the substrate and the corresponding EDX mapping of iron, revealing Fe presence solely in the upper part. Additional SEM and EDX data, including elemental fractions, are provided in Supporting Information S2. Although SEM and EDX analysis visually confirm particle distribution, these techniques are not directly quantitative.

Iron quantification was performed using two methods: a colorimetric assay via UV-Visible spectroscopy and Vibrating Sample Magnetometry (VSM). The first method involved measuring the mass absorption coefficient $\epsilon_m$ of iron in FeMPs and in magnetic pillars dissolved in 37% hydrochloric acid (HCl), comparing these values to the $\epsilon_m$ of iron with a reference. For this reference, iron data were obtained from 10 nm maghemite ($\gamma$-Fe$_2$O$_3$) nanoparticles dissolved in HCl, and then normalized by iron content.[46,47] In Fig. 1d, the mass absorption coefficient of FeMPs is shown to be reduced by 32% compared to pure iron, suggesting the microparticles contain in average 68% elemental iron (Fe$^0$), with the remainder likely due to carbonyl groups or additives. The curve in blue in Fig. 1d represents $\epsilon_m$ for 4067 magnetic pillars dissolved in concentrated HCl. Comparing this with the FeMP reference reveals an average volume fraction $\phi_{FeMP}$ of 22% in the pillars. Values for $h_{FeMP}$ and $\phi_{FeMP}$ are key parameters for subsequent pillar bending simulations.

Further iron quantification was performed using VSM. Magnetization curves $M(H)$ for both FeMPs powder samples and pillar arrays were measured as a function of the magnetic field (Supporting Information S3). For the powder, the data was normalized by the effective volume of FeMPs in the sample. On the other hand, the raw $M(H)$ pillar data was normalized by the total pillar volume within the investigated array. As shown in Fig. 1e, both systems display similar behavior: a linear increase at low fields, followed by a saturation plateau around $10^6$ A m$^{-1}$, with a lower saturation value for the pillars. The absence of hysteresis in the field cycles confirms superparamagnetic behavior. Analysis of the linear region suggests a magnetic particle size estimate of 3 nm, slightly smaller than the 9 nm-crystallite size measured by wide-angle X-ray scattering (Supporting Information S4). The FeMP saturation value $m_S$ = 1.90×10$^6$ A m$^{-1}$ is also in agreement with previous determination.[41] Comparing VSM data from Fig. 1e, the pillars reach a saturation magnetization of 20.5% relative to the FeMPs. This result aligns well with the colorimetric findings which indicated a value of 22%. Finally, this fabrication method achieves iron pillar concentrations that are twice higher than those obtained by directly mixing FeMPs with PDMS prior to molding,[41] with the added benefit of localizing iron in the upper part of the pillars.[39]

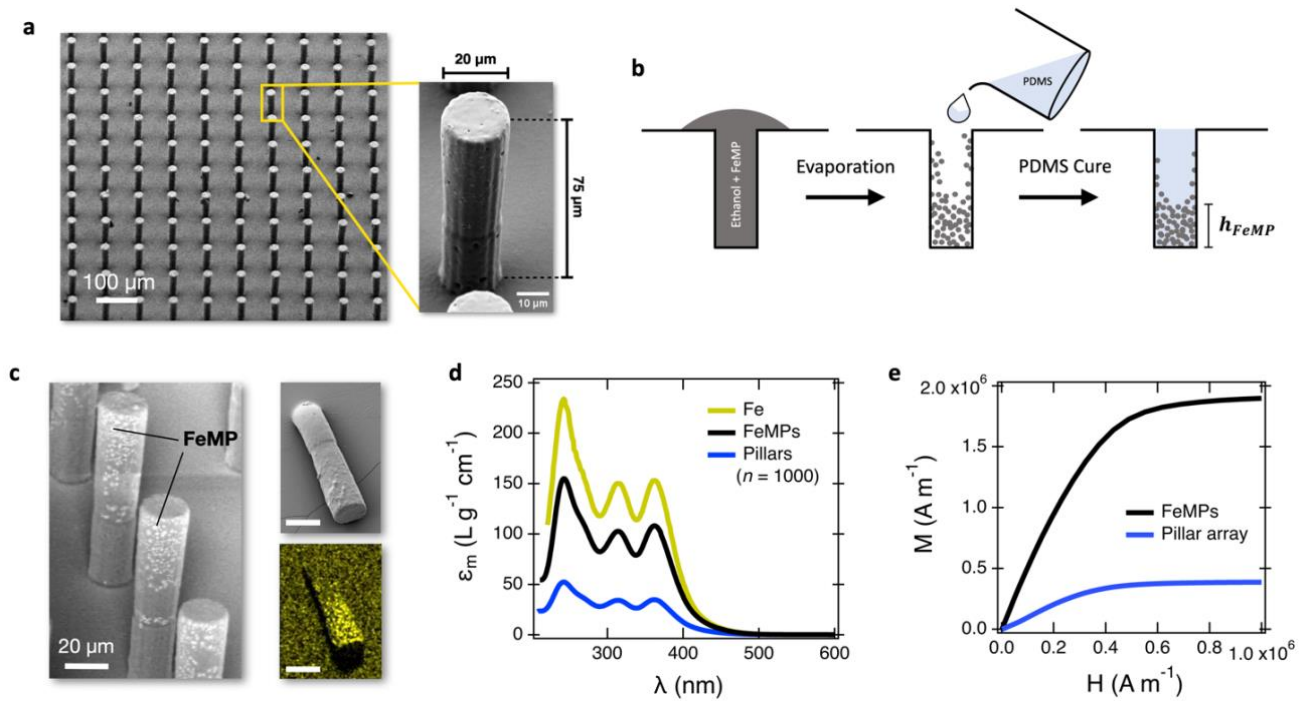

**Figure 1: a)** Scanning electron microscopy (SEM) image of a PDMS pillar array. Inset: close-up view of a single pillar highlighting its geometric dimensions ($d_{Pillar}$ = 20 µm, $l_{Pillar}$ = 75 µm). **b)** Schematic of the fabrication process for magnetic pillars: a slurry of ethanol and FeMPs is poured into the counter-mold; ethanol is then evaporated, and PDMS elastomer is added over the particles and cured. **c)** SEM image of a magnetic pillar array, showing FeMPs as white dots concentrated in the upper part of the pillars; insets: SEM image and iron EDX mapping of a pillar bent toward the substrate. **d)** Mass absorption coefficient, $\epsilon_m$, of pure iron, FeMPs, and magnetic pillars dissolved in high-concentration hydrochloric acid. The absorbance signal reflects the formation of $FeCl_4^-$ (tetrachloroferrate) ions, allowing for iron quantification.[46] **e)** Magnetization curves comparing FeMP powder and magnetic pillars.

## 2.2 - Magnetic Bending
### 2.2.1 - Experimental measurement of pillar deflection

The bending response of the pillars was characterized by exposing them to a magnetic field generated by two 12 mm-cubic N48 neodymium magnets terminated by an iron triangular prism (Fig. 2a). Using a micromanipulator, the magnet tip was moved along the $y$-axis and maintained at a fixed height of 200 µm above the pillars during approach. Fig. 2b shows the magnetic field $B(y)$ and the magnetic field gradient $dB(y)/dy$ as a function of the magnet distance to the microchip center. The maximum magnetic field, approximately 0.4 T, was recorded directly beneath the prism tip, corresponding to a gradient of 250 T m$^{-1}$. Notably, at distances greater than 10 mm from the magnet tip, the field gradient nears zero. As the magnet moves closer to the array, the pillars bend due to the concomitant influence of magnetic force and torque. Fig. 2c displays a chronophotograph of three neighboring pillars in the array at varying distances from the magnet tip, ranging from 12 mm to 0 mm, clearly illustrating that the deflection $\delta_{Pillar}(y)$ increases as the distance $y$ decreases (Supporting Information, movie#1). To quantify the deflection behavior, pillar bending was recorded by optical microscopy over 3 fields containing a total of 87 pillars in a sequence of images over time, then converted into distance data $\delta_{Pillar}(y)$. Fig. 2d presents these deflections as a shaded surface across all 87 pillars. The deflection pattern is consistent for all collected data, showing an inflection point around $y = 3$ mm and reaching a saturation as the magnet approaches directly above the pillars. Maximum deflection values for this array range from 35 to 45 μm. Moderate variability in pillar responses is observed, likely due to differences in FeMP distribution or pillar elasticity. The curve in red at the center of the shaded region represents the average deflection, calculated as a 10-point moving average. Below 1 mm, the deflection plateaus, which could be attributed to various factors, including the geometry of the magnetic field that increasingly aligns or pulls pillars upward toward the magnet as it nears. The following section presents finite element simulations of pillar bending, analyzing the effects of induced magnetic forces and torques.

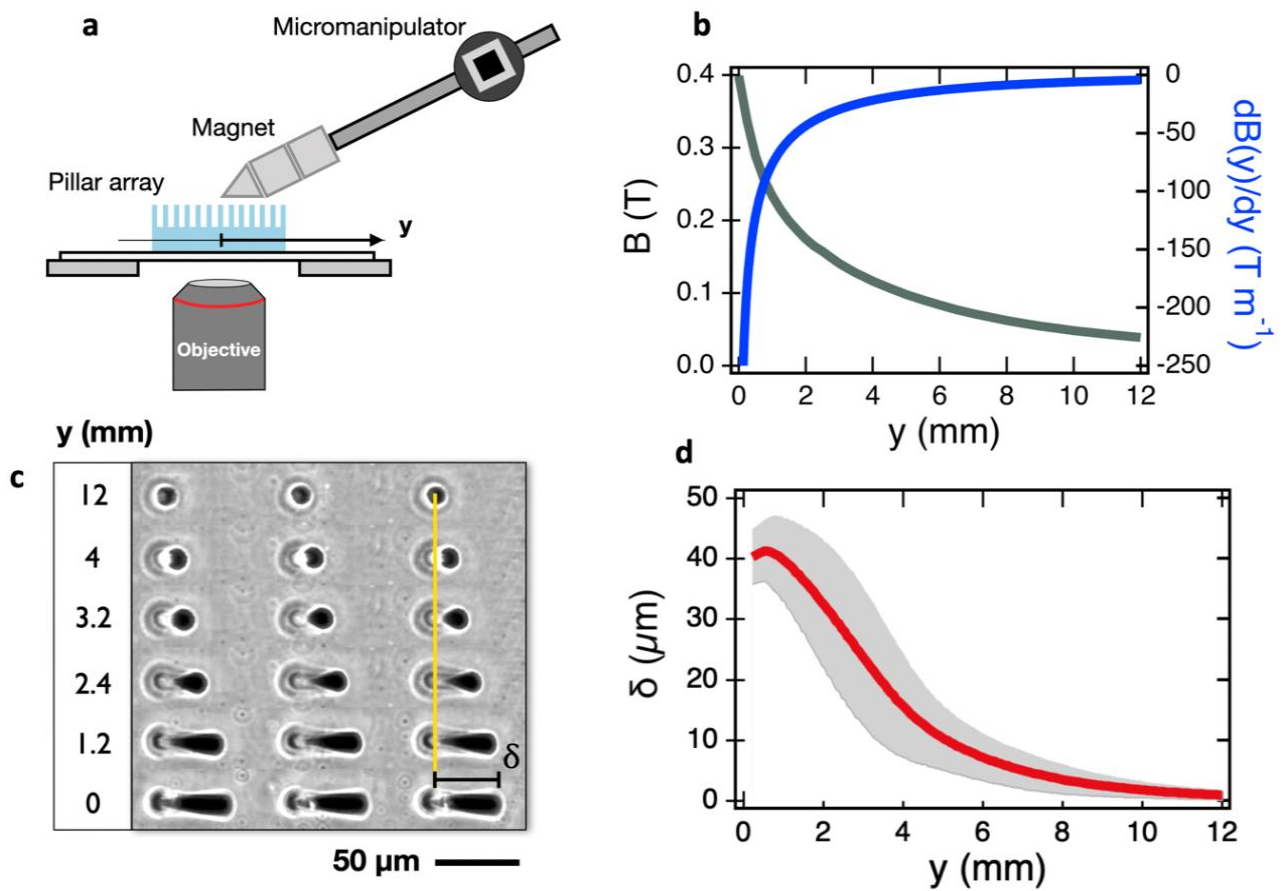

**Figure 2**: **a)** Schematic representation of the device used to study pillar bending featuring phase contrast optical microscopy for visualization. **b)** Magnetic field $B(y)$ and magnetic field gradient $dB(y)/dy$ generated by two 12 mm-cubic N48 neodymium magnets 200 μm above the pillar tips as a function of the distance $y$. **c)** Chronophotograph showing the bending of pillars as the magnet moves closer to the chip center. The deflection, $\delta_{Pillar}(y)$, is marked in the image. **d)** Variation of deflection $\delta_{Pillar}(y)$ as a function of the distance between pillars and magnet. The area in gray represents the response of 87 individual pillars, whereas the curve in red is the 10-point moving average (see also movie#1 in Supporting Information).

**2.2.2 - Comparison of Simulated and Experimental Bending**

To simulate the bending of the pillar, we first modeled the magnetic field generated by the neodymium magnets (Fig. 2a) using COMSOL 3D current-free magnetic field module.[48] The resulting magnetic field isolines, shown in Fig. 3a, indicate a rapid decrease in field strength with distance from the magnet, consistent with experimental measurements (Fig. 2b). Additionally, the isolines reveal that the field direction becomes increasingly horizontal with distance from the magnetic tip and nearly vertical directly beneath it.

Pillar deflection was simulated separately using the COMSOL beam simulation environment.[48] A single pillar, modeled as a slightly tapered beam 75 μm long and 20 μm in diameter, fixed at its base and free at its end, was designed to match SEM measurements. The beam was subjected to a magnetic force as well as a magnetic torque perpendicular to the bending plane. The bending of a simulated pillar for a magnet located at 10, 2, 1 and 0 mm is shown in Fig. 3b. As the magnet gets closer, bending increases. The simulations provide the Von Mises stress distribution along the beam, visualized by the color scale. It is found that the stress is highest at around 20 μm above its base. The Von Mises stress values indicate that the pillar is only slightly outside the region of linear behavior for typical PDMS.[49] This occurs when the pillar is just below the prism extremity.

Fig. 3c presents a direct comparison of pillar bending driven by magnetic force (in blue) alone *versus* combined magnetic force and torque (in shades of red) effects. In this simulation, the FeMP volume fraction is set to its experimental value, $\phi_{FeMP}$ = 22%, and the pillar Young modulus $E$ is fixed at 0.75 MPa. The simulation indicates that magnetic force alone underestimates the experimentally observed deflection (shaded region), highlighting that both torque and force contributions are essential for accurate interpretation. This represents a key finding of this work: magnetic force alone cannot account for the deflection of PDMS pillars under current conditions.

Simulations were also conducted across various $h_{FeMP}/l_{Pillar}$ values, with optimal agreement closely matching the experimental data in Fig. 1c. At short distances, deflection saturation occurs due to the magnetic field predominantly vertical orientation, which aligns the pillar vertically as well. As a second parameter, we varied the PDMS material properties to simulate the effect of elasticity on pillar bending (Fig. 3d), conducting simulations at $h_{FeMP}/l_{Pillar}$ = 0.5 and $\phi_{FeMP}$ = 22%. We found that increasing $E$ led to a decrease in deflection $\delta_{Pillar}$. The relationship between $E$ and $\delta_{Pillar}$ is however nonlinear, with the most pronounced effect occurring at intermediate distances (1–5 mm). Compared to experimental data, the stiffest and softest beams fell outside the 95% confidence interval, with the best fit observed for $E$ = 0.75 MPa.

In a final simulation, we examine the effect of the magnetic field on pillar deflection (Fig. 3e), keeping the parameters $E$, $h_{FeMP}$ and $\phi_{FeMP}$ as before. Moreover, the magnetic field strength was modeled based on the experimental data shown in Fig. 2b. As $B$ increases, the simulated pillar deflection also rises, consistent with the quadratic dependence of torque on field amplitude. Fig. 3e further explores the impact of varying $\phi_{FeMP}$ on $\delta_{Pillar}$: reducing the FeMP fraction below the experimental value results in a decrease in deflection, with only $\phi_{FeMP}$ > 17% falling within the 95% confidence interval of experimental data. In summary to this part, finite element simulations show strong agreement with experimental results for various parameters, including $h_{FeMP}/l_{Pillar}$, $E$, and $\phi_{FeMP}$, validating the simulation scheme adopted here and underscoring the dominant role of magnetic torque in pillar deflection.

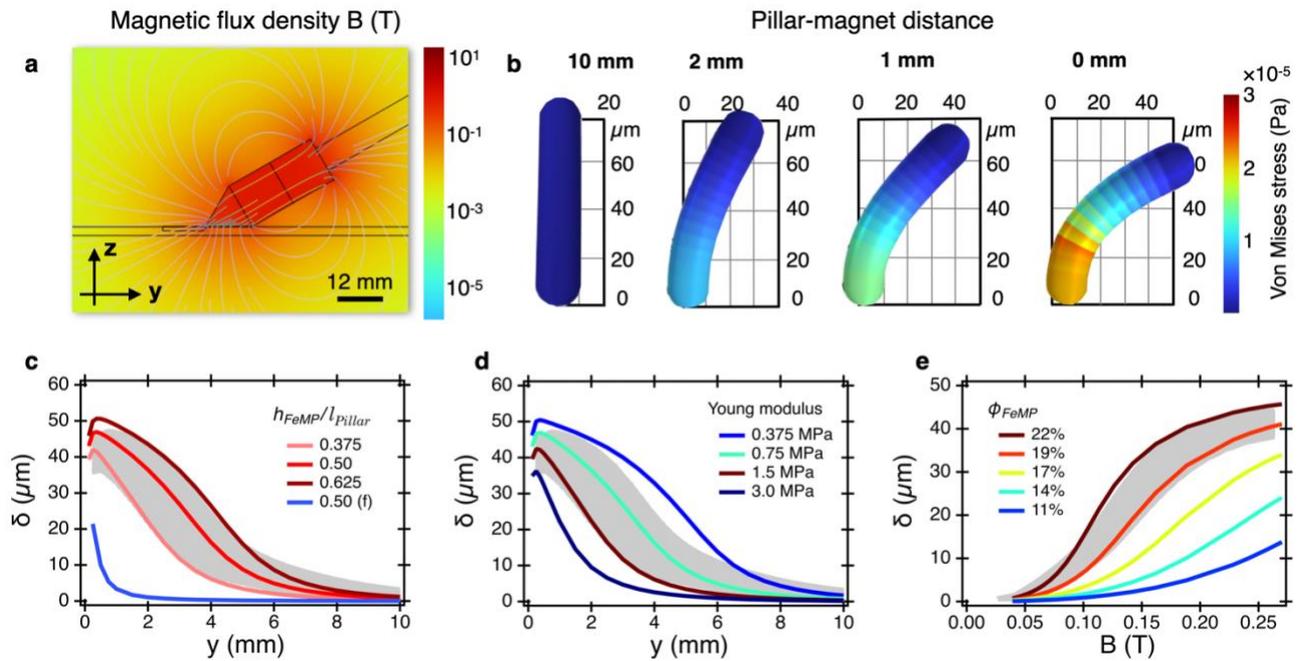

**Figure 3**: **a)** Simulated 3D-mapping of the magnetic field lines created by the magnetic pen and iron prism displayed in Fig. 2a. The field mapping was validated by comparing it with the field shown in Fig. 2b. **b)** The bending of a single pillar at distances $y$ = 10, 2, 1 and 0 mm (from left to right panels) from the magnet displayed in Fig. 3a. The pillar is modeled as a 75 µm-long, slightly tapered beam fixed at its base and free at its end. **c)** Simulated deflection $\delta_{Pillar}(y)$ as a function of the distance from the tip for different values of $h_{FeMP}/l_{Pillar}$, where $h_{FeMP}$ designates the height of the FeMP-loaded region and $l_{Pillar}$ the total pillar length (75 µm). In this simulation, $\phi_{FeMP}$ = 22%, and $E$ = 0.75 MPa. The curve in blue labelled "0.50 (f)" is obtained for force alone, while the curves in shades of red include force and magnetic torque. The area in gray represents the response of 87 individual pillars. **d)** Same as in Fig. 3c) simulating the effect of the pillar Young modulus $E$. **e)** Simulated deflection $\delta_{Pillar}$ as a function of the magnetic field $B$ for different values of the FeMP volume fraction $\phi_{FeMP}$.

## 2.3 - Closed-loop microchannel with pillar-driven flow
### 2.3.1 – Pillar actuation
We now build on the findings from the previous sections to design a circular closed-loop microchannel capable of generating a coherent flow (Figs. 4a-b). The microchannel incorporates a 150 µm-pitched array of 75 µm-high magnetic pillars, actuated by rotating neodymium magnets placed underneath to generate controlled flow within a $h$ = 275 µm deep channel. Real-time in situ observation are carried out *via* an optical microscope. A movie showcasing pillar movement during a typical experiment is provided in the Supporting Information (movie#2). Key stages of pillar orientation are illustrated in Figs. 4c-e: starting from an upright position, the pillars are gradually drawn rightward along a slightly off-center trajectory and eventually reach an intermediate tilt position. Fig. 4f presents the tip displacement of the pillar $x_{Pillar}(t)$ along the $x$-axis over two full magnet rotations, with rotation angles $\pi$, $2\pi$, $3\pi$ and $4\pi$ indicated at the top. The time points corresponding to Fig. 4c-e are also marked. The displacement profile exhibits asymmetric square-wave patterns, characterized by a gradual rise followed by a sharp drop. This asymmetry is even more evident in Fig. 4g, which shows the $x$-component of the tip velocity $dx_{Pillar}(t)/dt$. For much of the cycle, the velocity remains in the 100 µm s$^{-1}$ range, punctuated by spikes reaching velocities up to 20 mm s$^{-1}$ as the pillar snaps back to its equilibrium position. This time-dependent deflection closely resembles the behavior of lung cilia, with alternating recovery strokes and rapid whip-like movements. The whip-like motion is marked by sharp drops in $x_{Pillar}(t)$ at t = 0.175, 0.35, 0.525 and 0.70 s. Finally, Figs. 4f and 4g reveal distinct profiles between the 0/π and π/2π periods, attributed to slight misalignments in magnet positioning during the latter half of each rotation. However, this discrepancy has no significant impact on the system pumping efficiency.

In Fig. 4h, we examine the velocity spikes associated with the rapid whipping motion and its propagation along a row of pillars, shown here for a sequence of five pillars. The pillars, spaced 212 µm apart in the chosen configuration, exhibit a noticeable time shift of the velocity spikes, indicating a phase difference between their motions. This observation suggests the occurrence of a progressive wave in the pillar beating along the $x$-axis. By plotting the temporal phase shift against the pillar positions (inset), we obtain a linear relationship, where the slope corresponds to the inverse phase velocity $V_\phi$ = 48 mm s$^{-1}$, in good agreement with the magnet rotation speed. The progressive wave is also characterized by a wave vector $k$ = 0.73 mm$^{-1}$ and a wavelength $\lambda$ = 8.6 mm. To conclude this section, we have shown that in a closed-loop microfluidic chip with magnetic pillars standing on the bottom, a progressive wave in pillar beating is induced, exhibiting characteristics similar to the ciliary beat observed in certain organisms and organs.

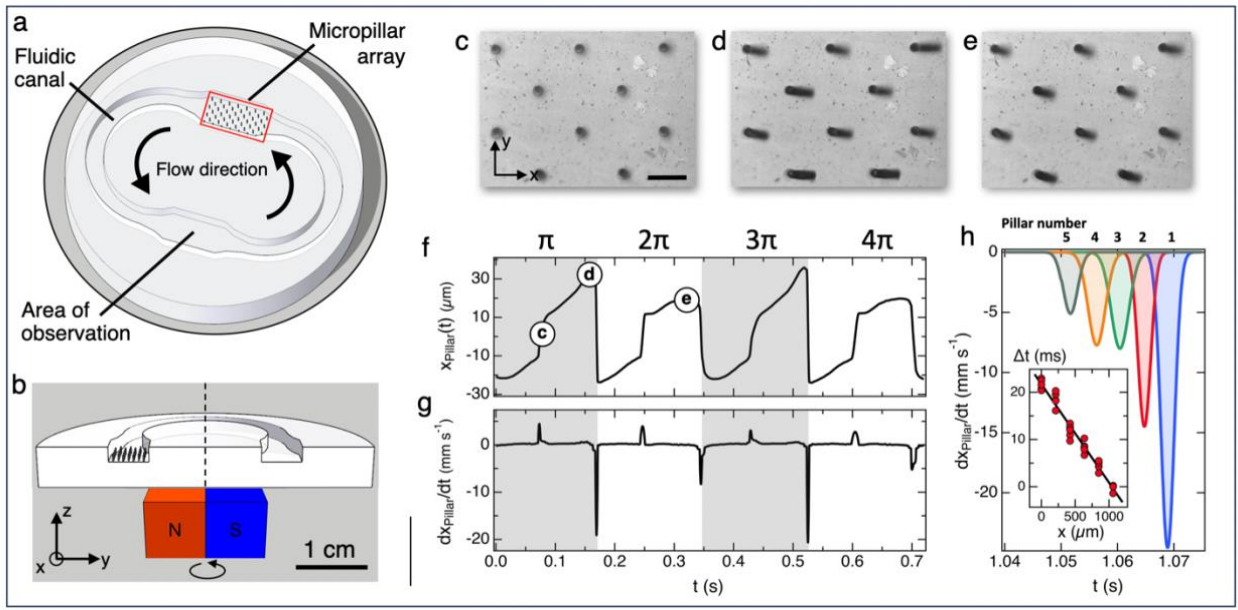

**Figure 4**: **a)** Top view schematic representation of the circular closed-loop microchannel, highlighting the pillar array, the fluidic canal and observation area. **b)** Side-view schematic of the setup in a), illustrating the placement of rotating magnets beneath the chip. **c, d, e)** Images of the pillars at various stages of their rotation cycle. The scale bar in panel c) represents 100 µm. **f)** Position of a pillar tip $x_{Pillar}(t)$ over two complete rotations of the magnet. **g)** Pillar tip velocity $dx_{Pillar}(t)/dt$ corresponding to the coordinates in figure f). **h)** Velocity profiles $dx_{Pillar}(t)/dt$ of five pillars aligned along the $x$-axis during the recovery stroke, showing a phase shift between pillars and indicating wave propagation.

**2.3.2 – Actuated Flow in Fluids of Increasing Viscosity**

To evaluate the pumping performance of actuating pillars, the microchannel was filled with fluid containing 2 µm-tracer beads, and sealed with a glass coverslip (Fig. 5a). The microscope focal plane was adjusted in 25 µm steps to reconstruct the flow profile within the channel. At each height, a time-lapse sequence was recorded at rest to ensure the channel was properly sealed, followed by one with the fluid being actuated by the pillar beating. The movement of the fluid was inferred by particle tracking velocimetry. A movie provided in the Supplementary Information (movie#3) shows that fluid displacement is reversed by changing the actuation direction. Furthermore, the flow within the channel halts almost immediately when the pillars cease motion, consistent with the low Reynolds number characteristic of microfluidic systems. Zooming in on particle displacement induced by pillar beating also reveals a pulsatile pattern, as illustrated in Supporting Information S5 and movie#4. It is found that the bead trajectories follow a sinusoidal function of the form: $x_{Bead}(t,z) = [dx_{Bead}(t,z)/dt]t + a(z)sin(2\omega t)$, where $x_{Bead}(t,z)$ and $dx_{Bead}(t,z)/dt$ are the bead coordinate and velocity along the $x$-axis respectively, $a(z)$ a height-dependent constant and $\omega$ the magnet angular frequency. This motion can be approximated simply by the linear term, yielding the velocity profile $V_x(z)$ after averaging over 10-100 beads.

Fig. 5b presents the flow profile of water at 21°C at the center of the observation zone. The experiment was performed using three different microchips, noted MC#1, MC#2 and MC#3 to evaluate the performance variability arising from fabrication. In all three cases, the maximum velocity $V_{Max}$ reaches values on the order of 10-15 µm s$^{-1}$, corresponding to Reynolds numbers $Re$ ~ 10$^{-3}$. The velocity maxima are located around 125-150 µm, i.e. roughly at the microchannel midplane. Regarding $V_{Max}$, we confirmed previous findings[20,38] that the maximum flow velocity is proportional to the magnet angular frequency $\omega$, and found $V_{Max}(\omega) = 0.84\omega$, where $V_{Max}$ is expressed in µm s$^{-1}$. The flow profile in Fig. 5b resembles Poiseuille flow in a Hele-Shaw channel but deviates slightly from the theoretical expression[50] $V_x(z) \sim z(h-z)$. Finally, it is worth noting that this flow differs from that found by Shields et al., for which the maximum was located at the pillar tips, and decreased above.[19]

Secondly, we studied velocity profiles in cetylpyridinium chloride (CPCl) and sodium salicylate (NaSal) viscoelastic wormlike micellar solutions (Figs. 5c-f). In the semi-dilute concentration regime ($c_{CPCl-NaSal}$ > 0.3 wt. %) these micelles form an entangled network similar to that of polymers.[51] For CPCl-NaSal concentrations between 0.5 and 4 wt.%, the static viscosity $\eta_0$ increases from 0.007 to 20 Pa s and the elastic moduli $G_0$ from 0.2 to 20 Pa (Table I), covering the typical viscosity and elasticity ranges of pulmonary mucus.[52-54] The velocity profiles obtained with microchannel MC#2 exhibit a Poiseuille-like flow, similar to that of water, with velocity maxima slightly off-center. As surfactant concentration—and consequently viscosity—increases, maximum velocities decrease sharply. To quantify this effect, we have plotted in Figs. 5g-i the flow rate $Q$, the maximum shear rate $\dot{\gamma}_{Max}$ and shear stress $\sigma_{Max} = \eta_0 \dot{\gamma}_{Max}$ generated by pillar actuation as a function of $\eta_0$, respectively. For both $Q$ and $\dot{\gamma}_{Max}$, a sharp decrease is observed at low viscosity, followed by a plateau for $\eta_0 > 1$ Pa s. Conversely, $\sigma_{Max}$ is found to be stable at low viscosity and then increases linearly with $\eta_0$. The derived $\dot{\gamma}_{Max}$ for wormlike micelles are notably small, on the order of $10^{-2}$ s$^{-1}$, leading to a Weissenberg number $W_i = \dot{\gamma}_{Max}\tau \ll 1$, where $\tau$ is the micellar network relaxation time (Table I).[51,55,56] In rheology, particularly for viscoelastic liquids, the inequality $W_i \ll 1$ unequivocally indicates that the flow occurs in the Newtonian regime, where viscosity *versus* shear rate remains constant.[56] This indicates that, on average, the flow of surfactant solutions remains that of a Newtonian fluid across the chip height.

We also observed that pillar beating remains unaffected by fluid viscosity; for viscosities ranging from 0.001 to 20 Pa·s, the displacement $x_{Pillar}(t)$ and velocity $dx_{Pillar}(t)/dt$ profiles are consistent with those shown in Figs. 4f-4g. This similarity indicates that the chip operates as a stress-controlled rheometer, applying a stress of approximately of $10^{-4}$ Pa (Fig. 5i). As viscosity increases, the relationship $\dot{\gamma}_{Max} = \sigma_{Max}/\eta_0$ holds, leading to a maximum shear rate that scales inversely with viscosity ($\eta_0^{-1}$), as demonstrated by the straight line in Fig. 5g-h. Locally however, particularly near the pillars, velocities can reach up to 20 mm s$^{-1}$ during their whip-like motion, corresponding to angular frequencies of $\omega_{Pillar}$ = 250 rad s$^{-1}$. Under these conditions, the product $\omega_{Pillar}\tau$ is much larger than one, demonstrating that during this short period, surfactant solutions with concentrations higher than 1 wt. % behave as a purely elastic material. We propose that the observed deviations in $Q(\eta_0)$ and $\dot{\gamma}_{Max}(\eta_0)$ from the expected $\eta_0^{-1}$-behavior arise from the elasticity of the micellar network. These findings will need to be validated using data from mucus gel samples.

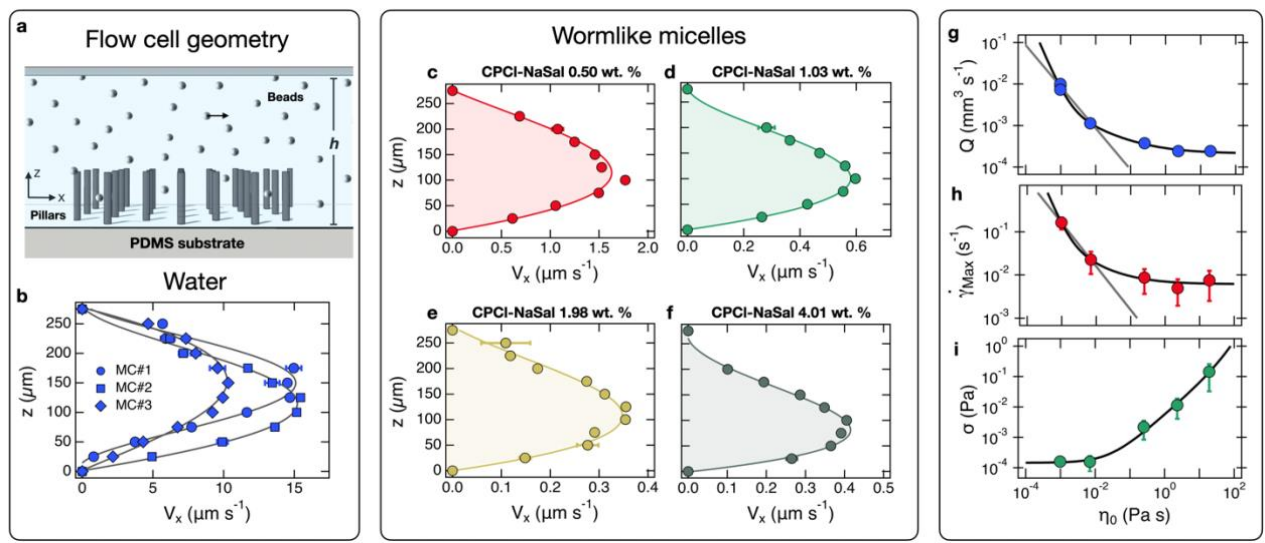

**Figure 5**: **a)** Schematic of the actuation area and flow cell geometry with height $h$ = 275 μm. **b)** Flow velocity profiles across the observation zone using water at 21°C for chips MC#1, MC#2, and MC#3, confirming reproducibility. **c-f)** Velocity profiles for CPCl-NaSal viscoelastic solutions at concentrations $c_{CPCl-NaSal}$ = 0.50, 1.03, 1.98 and 4.01 wt.%, corresponding to static shear viscosities $\eta_0$ of 0.007, 0.25, 2.3 and 19.1 Pa s at 21°C; Inset: representation of CPCl-NaSal wormlike micelles in the semi-dilute regime. **g-i)** Viscosity dependences of the flow rate $Q$ (g), maximum shear rate $\dot{\gamma}_{Max}$ (h) and maximum shear stress $\sigma_{Max}$ (i) generated by the pillar actuation in microchannel

MC#2. In g) and h), the straight line corresponds to the behavior $\dot{\gamma}_{Max} \sim \eta_0^{-1}$. The continuous line are guides for the eyes.

# 3 – Conclusion

This study presents the development and characterization of magnetically actuated flexible PDMS pillar arrays as a tool for microfluidic flow generation and analysis. Fabricated using soft lithography, the pillars were embedded with iron carbonyl microparticles, allowing for precise magnetic responsiveness. Structural characterization, including SEM and EDX analysis, revealed uniform FeMP distribution, predominantly localized in the upper half of the pillars, a key factor in achieving controlled bending behavior. Quantitative magnetic and mechanical analyses demonstrated the critical contributions of both magnetic force and torque in pillar deflection, as confirmed through experiments and finite element simulations. The integration of these pillars into a closed-loop microchannel enabled effective flow actuation in low Reynolds number conditions, mimicking biological ciliary motion. Through systematic flow measurements, the study demonstrated consistent Poiseuille-like flow profiles in Newtonian fluids and explored the impact of increasing fluid viscosity and viscoelasticity, achieving stable actuation across a wide viscosity range. Key findings include the scaling of maximum shear stress with fluid viscosity and the emergence of plateau behaviors in flow rate and shear rate, indicating a regime transition between viscous and viscoelastic flows. The observed flow patterns, combined with the controllability and versatility of this platform, highlight its potential for studying biological fluid transport, mimicking pulmonary mucus clearance, or developing microfluidic systems for complex fluid handling. Future work may focus on refining pillar designs, expanding flow applications, and integrating this platform with real-time sensing technologies for more advanced biological and industrial applications.

# 4 – Materials and Methods
## 4.1 – Materials
### 4.1.1 – Chemical compounds

For microchip fabrication, undoped 7.6 cm silicon wafers (thickness 380 µm) were purchased from Neyco Vacuum & Material (Vanves, France), and polydimethylsiloxane (PDMS) Sylgard™ 184 base and crosslinkers from Dow Chemical (Midland, USA). Dry film photosensitive resin layers, with thicknesses of 25 µm, 50 µm, and 100 µm, were acquired from Nagase ChemTex (Osaka, Japan), and cyclohexanone and trichloro(1H,1H,2H,2H-perfluorooctyl)silane from Sigma-Aldrich (Saint-Quentin-Fallavier, France). Iron carbonyl microparticles (FeMPs) used as a PDMS additives were from Neochimie (Cergy-Pontoise, France). Wormlike micellar solutions were prepared using cetylpyridinium chloride (CPCl), sodium salicylate (NaSal), and sodium chloride (NaCl) from Sigma-Aldrich (Saint-Quentin-Fallavier, France). For particle tracking experiments, Polybead® microspheres (diameter 2.00 µm) were purchased from Polysciences Inc. (Warrington, USA).

### 4.1.2 – Fabrication of pillar array and microchannel

The pillar array (Fig. 1a) and microchannel (Fig. 4a) were made using PDMS-based microlithography and Klayout viewer and editor (https://www.klayout.de/). For the mold template, dry film negative photoresist layers were laminated to an adhesion-treated silicon wafer. The laminated wafer was then placed in a maskless UV lithography MicroWriter ML3 (Cambridge UK), and the pillar and channel geometry were selectively exposed to UV light (intensity 1.05 J cm$^{-2}$). When multiple design layers were needed, the lamination and exposure processes were repeated for a second design after a post-exposure bake. The exposed wafer and resin were then developed in cyclohexanone, before a final hard bake to fully harden the mold. The master template was silanized after oxygen plasma treatment and exposure to vaporized trichloro(1H,1H,2H,2H-perfluorooctyl)silane to prevent adhesion, then coated with PDMS (10:1 base-to-crosslinker ratio). The PDMS counter-mold was baked at 150 °C for 60 min to ensure complete crosslinking.

Magnetic pillars were fabricated first by filling the pillar wells with FeMPs. The design consisted of a 1×1 cm square containing an array of pillars with diameters of 20 µm, arranged on square pitches of 66 or 150 µm, and reaching a height of 75 µm. A 50 wt. % solution of FeMP in ethanol was pipetted over the counter-mold at a coverage of 2 µL per square millimeter of the pillar array. This FeMP slurry was then rubbed into the mounter-mold surface containing pillar wells. The excess ethanol was removed by placing the counter-mold in a vacuum chamber, leaving only dry FeMPs. The excess FeMP on the counter-mold surface was removed using microfiber damp with ethanol. The filling process described above was repeated 5 times. The subsequent steps for fabricating the array follow the same procedure outlined earlier for filling the counter-mold with PDMS.

For the microchannel design, the first layer consisted of both the microchannel walls as well as a 4×6 mm pillar array with circles measuring 20 µm in diameter, spaced 150 µm apart on a square lattice (Fig. 4a). The first layer was laminated with a 25 µm photoresist layer, followed by a 50 µm layer before UV exposure (leading to $l_{Pillar}$ = 75 µm), while the second layer consisted solely of the walls. Two additional 100 µm photoresist layers were laminated and exposed to UV solely on the microchannel walls, resulting in a total height of 275 µm. Care was taken to align the two layers of the design using the maskless lithography system. The pillar array is positioned 7.5 mm from the center of the channel loop, opposite to a pillar free observation area and linked by a fluidic canal. To actuate the magnetic pillars, two neodymium magnets were positioned beneath the microchannel, as illustrated in Fig. 4b. These magnets were mounted on a custom stand connected to an electric motor, which enabled controlled rotation in either direction at frequencies $\nu$ ranging from 0 to 2.8 Hz. Their alignment ensured that the pillars alternated between a strong and weak magnetic field during rotation. The entire setup was mounted on an upright optical microscope with a 10× objective to facilitate real-time observation of pillar movement. Flow profiles measurements in water were conducted using three different microchips, labeled MC#1, MC#2, and MC#3, to assess performance variability due to fabrication.

### 4.1.3 – Wormlike micellar solutions

The wormlike micelle solutions were obtained by mixing cetylpyridinium chloride surfactant (CPCl) and sodium salicylate (NaSal) hydrotrope in a 0.5 M NaCl brine.[51,55] The total surfactant concentration $c_{CPCl-NaSal}$ was varied between 0.5 and 4 wt.%, while the [NaSal]/[CPCl] molar ratio was kept constant at 0.5.[51,55] Under these conditions, CPCl and NaSal spontaneously self-assemble into long micron-sized wormlike micelles that behave as Maxwell fluids, i.e. the stress relaxation function decreases like an exponential with a unique relaxation time.[57] Table I summarizes the values of viscosity $\eta_0$, elastic modulus $G_0$ and relaxation time $\tau$ for the 4 concentrations studied.[51,55] For the lowest concentrations, elastic modulus values were obtained from the ratio $\eta_0/\tau$.

| $c_{CPCl-NaSal}$ | $\eta_0$ | $G_0$ | $\tau$ |
|---|---|---|---|
| wt. % | Pa s | Pa | s |

| | | | | |
|---|---|---|---|---|
| Water | 0 | 0.00098 | n.d. | n.d. |
| CPCl-NaSal | 0.50 | 0.007 | 0.17 | 0.04 |
| CPCl-NaSal | 1.03 | 0.25 | 1.5 | 0.17 |
| CPCl-NaSal | 1.98 | 2.3 | 7 | 0.34 |
| CPCl-NaSal | 4.01 | 19.1 | 29 | 0.71 |

**Table I** Rheological characteristics of water and wormlike micellar solutions made from cetylpyridinium chloride and sodium salicylate, at [NaSal]/[CPCl] = 0.5 under the experimental conditions used (T = 21°C). The data on viscoelastic liquids are in good agreement with those originally published.[51,55]

## 4.2 - Methods

### 4.2.1 – Quantification of iron content in magnetic pillar array

The iron microparticle content in the pillars was quantified using an iron titration protocol from previous studies.[46,58] A magnetic pillar array with a known pillar number was placed in a hydrochloric acid solution (37%), which effectively dissolves the embedded FeMPs. The number of pillars per array, varying from 500 to 5000 was determined based on images captured on a binocular microscope before the dissolution (Supporting Information S6). Dissolution was confirmed by the yellow color obtained within a few minutes,[46] a distinctive feature of tetrachloroferrate ions $FeCl_4^-$. The absorbance of the resulting solution $A_{Pillar}(\lambda)$ was measured using UV-Visible spectrophotometry.

### 4.2.2 - Pillar Bending by magnet translation

Two 12 mm-cubic N48 neodymium magnets with an iron focusing prism were mounted on a micromanipulator (M-LSM, Zaber Technologies) to enable horizontal movement along the $y$-axis (Fig. 2a). The magnetic pen was positioned 200 µm above the pillars at a 30° angle from horizontal. Magnetic field measurements were taken at the top of the pillars using a gaussmeter (GM08, Hirst Magnetic Instruments). Pillar arrays were observed by phase contrast optical microscopy and were recorded as the magnet approached at a speed at 125 µm s$^{-1}$.

### 4.2.3 – Pillar Bending by magnet rotation

The microchannel was placed on top of a plain silicon wafer and its center was aligned with the center of the rotating magnetic axis. The magnetic axis was designed such that a motor would rotate a magnet holder for two 12 mm cubic neodymium magnets. The magnetic axis mount was affixed to the microscope stage, and had its position adjusted by thumb screws. The motor was controlled by an external controller hub, which was capable of rotating the motor at up to 2.8 Hz in either direction. Pillar actuation and flow tests were performed using an upright microscope equipped with a 10x objective.

### 4.2.4 - Magnetic field and beam simulations

The magnetic field generated by a magnetic pen was modeled using COMSOL Multiphysics Version 6.2 (license number 8079561), simulating the field without electrical currents. The model uses a 20 mm cubic simulation space with magnetic insulation on all boundaries, excluding the symmetry plane. To simulate the bending of a magnetic pillar, the study utilizes a beam model. The pillar is represented as a circular beam with a diameter varying linearly from 19 µm at its base to 21 µm at its tip over its 75 µm length, and is defined with uniform elastic modulus $E$ = 750 kPa. External magnetic forces are applied along the free tip of the beam, balancing forces by solving beam equations.[60] The magnetic force acting upon the beam is calculated along the magnetized beam as:

$$\boldsymbol{F_{Mag}} = (\boldsymbol{M} \cdot \boldsymbol{\nabla})\boldsymbol{B} \qquad (1)$$

Meanwhile, the torque acting along the same beam length is obtained from:[61,62]

$$\boldsymbol{\Gamma_{Mag}} = \frac{\mu_0 B^2 M^2 (N_{YY} - N_{ZZ})\boldsymbol{e_l} \cdot \boldsymbol{e_B}}{(B + \mu_0 M(N_{ZZ} - 1))(B + \mu_0 M(N_{YY} - 1))} \boldsymbol{e_l} \times \boldsymbol{e_B} \qquad (2)$$

For simulations, magnetization data is taken from the measured magnetization curve shown in Fig. 1c such that $M(B) = \phi M_{FeMP}(B)/\phi_{max}$, noting that in a vacuum, $B = \mu_0 H$. The shape demagnetization, $N_{ZZ}$ and $N_{YY}$ in Eq. 2 correspond to that of a cylinder aligned in $z$-direction.[62] Finally, the direction vectors $\boldsymbol{e_l}$ and $\boldsymbol{e_B}$ correspond to the direction of the beam and $\boldsymbol{B}$ respectively. The simulation captures changes in the magnetic field gradient over time, which impacts the pillar bending response, as the applied field varies with the position of the magnetic pen in the simulated environment. Full details of the calculations for these simulations are provided in Supporting Information S7.

**4.2.5 – Vibrating Sample Magnetometry (VSM)**

The magnetization of magnetic pillar arrays was measured by VSM (Quantum Design). The pillar arrays were stuck onto nonmagnetic holders and mounted in the magnetometer. Measurements were performed at room temperature (T = 25 °C), using a frequency of 40 Hz. Cycles of magnetic excitation increasing from 0 to 1.59×10$^6$ A m$^{-1}$ and then decreasing back to 0 were performed at a scanning speed of 8×10$^3$ A m$^{-1}$ s$^{-1}$. The number of pillars per sample was determined based on images captured on a binocular microscope before the magnetization measurement. The magnetization of FeMP was measured under the same field and frequency conditions as above.

**4.2.6 - Optical microscopy and particle tracking**

Pillar bending by magnet translation was measured using an IX73 Olympus inverted microscope equipped with a 10X objective (numerical aperture 0.30). Time-lapse images during magnet translation (Fig. 2) were acquired in phase contrast via an EXi Blue camera (QImaging) and Metaview software (Universal Imaging Inc.). Pillar deflection was measured by tracking the pillar top position using the TrackMate plugin in ImageJ.[63] Pillar rotational actuation and flow tests (Figs. 4 and 5) were performed using an Olympus BX51 fluorescence microscope with a 10 × objective (numerical aperture 0.30) in reflected bright field. To determine the velocity field in the microchannel, 2 µm polystyrene beads (Polys-ciences Inc., Warrington USA) were suspended at a concentration of 0.01 vol. % in DI water and CPCl-NaSal wormlike micelles. Movies for Particle Tracking Velocimetry (PTV) were captured with a Fujifilm XE1 camera at the frame rate of 30 fps, while high-speed recordings, such as those of pillar actuation, were made using a RIBCAGE-modified Sony RX0 II camera set at 240 fps. Image sequences were analyzed using the TrackMate[59] particle tracking plugin operated within the ImageJ software platform.[63] Track data was then fitted to estimate diffusion, drift and mean horizontal and vertical linear motion.

4.2.7 – UV-Visible spectrometry

The absorbance of hydrochloric acid solutions containing FeMPs was analyzed using a UV-visible spectrophotometer (JASCO, V-630) with a temperature control feature. The absorption spectra $A_{Fe}(\lambda)$ were recorded from 190 to 800 nm at room temperature (25 °C). Based on the Beer-Lambert law, the relationship between $A_{Fe}(\lambda)$ and the iron concentration $c_{Fe}$ is expressed as $A_{Fe}(\lambda) = \varepsilon_{Fe}(\lambda) l c_{Fe}$ where $\varepsilon_{Fe}(\lambda)$ represents the absorptivity coefficient and $l$ is the cuvette thickness. $\varepsilon_{Fe}(\lambda)$ s then converted to the mass absorption coefficient $\epsilon_m$ of the iron contained in FeMPs and in magnetic pillars.

4.2.8 – Linear shear rheology
The complex elastic modulus $G^*(\omega) = G'(\omega) + iG''(\omega)$, where $G'(\omega)$ and $G''(\omega)$ denote the storage and loss moduli, was obtained using an ARES-G2 rheometer (TA Instruments) working with a cone-and-plate geometry (diameter 50 mm, cone angle 0.04 rad, sample volume 1.5 mL, gap 0.0555 mm). Experiments were carried out on cetylpyridinium chloride and sodium salicylate wormlike micellar solutions at total concentration $c_{CPCl-NaSal}$ = 0.50, 1.03, 1.98 and 4.01 wt. %, and temperatures of 21°C and 37 °C. Strain sweep tests were performed at $\omega$ = 1 rad s$^{-1}$ with a strain increasing from 1 to 100% and frequency sweeps from 0.1 to 100 rad s$^{-1}$. For frequency sweep experiments, the strain was kept constant at 10%, except for micellar solutions with concentrations of 0.50 and 1.03 wt. %, where the strain was increased to 100% and 50%, respectively. The data for $G'(\omega)$ and $G''(\omega)$ were analyzed using the Maxwell model, allowing us to derive the static shear viscosity $\eta_0$, the instantaneous shear modulus $G_0$, and the relaxation time $\tau$ of the surfactant solutions (Table I).

4.2.9 – Scanning electron microscopy (SEM) and Energy-dispersive X-ray spectroscopy (EDX)
A series of SEM scans, including EDX measurements, was carried out after metallization of the samples with a 15 nm platinum film to identify the presence of iron in the pillar arrays. SEM and EDX analyses were conducted on a GeminiSEM 360 microscope (Zeiss) operating at 5 or 10 kV and equipped with an Ultim Max detector (Oxford Instruments, 170 mm$^2$). The SEM images and EDX mapping were obtained by SE2 or InLens secondary electron detectors. Oxford Instruments AZtec software was used to acquire EDX maps and local composition analyses.

# Author contributions
Conceptualization: JF, JFB, NT, SL – Methodology: CPM, JF, JFB, MLG, SR – Software: CPM – Formal analysis: CPM, JF, JFB, MLG, SR – Investigation: CPM, JF, JFB, MLG, SR – Supervision: JF, JFB, SL – Writing – original draft: CPM, JF, JFB – Writing – review and editing: JFB – Funding acquisition: JF, JFB, NT, SL

# Data availability statement
The data supporting this article have been included as part of the Supplementary Information.

# Supporting Information
Size distribution of the iron carbonyl microparticles (S1) – Elemental maps and fractions of carbon, silicon, iron and oxygen obtained from energy-dispersive X-ray (EDX) analysis (S2) – Iron carbonyl microparticle (FeMPs) powder magnetization curve (S3) – Wide-Angle X-Ray Scattering (WAXS) iron microparticles (S4) – Particle trajectory oscillations and corresponding fitting (S5) – Image of magnetic micropillar arrays in binocular microscopy (S6) – Detailed simulation procedure (S7)

# Acknowledgments

We acknowledge the ImagoSeine facility (Jacques Monod Institute, Paris, France), and the France BioImaging infrastructure supported by the French National Research Agency (ANR-10-INSB-04, « Investments for the future »). We also thank the ITODYS SEM facility (Université Paris Cité, CNRS UMR 7086, Paris, France) and Dr Sarra Derouich for her help with the SEM and EDX analyses. David Montero from the IPM SEM facility (Sorbonne Université, FR2482, Paris, France) is also acknowledged for help with the secondary SEM analysis. We thank David Hrabovsky from the MPBT platform (Sorbonne University, FR 2769, Paris, France) for VSM measurements. We would like to acknowledge the work of Frederik Gélébart from the PHENIX laboratory (Sorbonne Université, UMR 8232, Paris France) for designing and assembling the motorized magnetic axis which was affixed to our microscope. This research was supported in by the Agence Nationale de la Recherche under the contract ANR-17-CE09-0017 (AlveolusMimics) and ANR-21-CE19-0058-1 (MucOnChip).

# TOC image

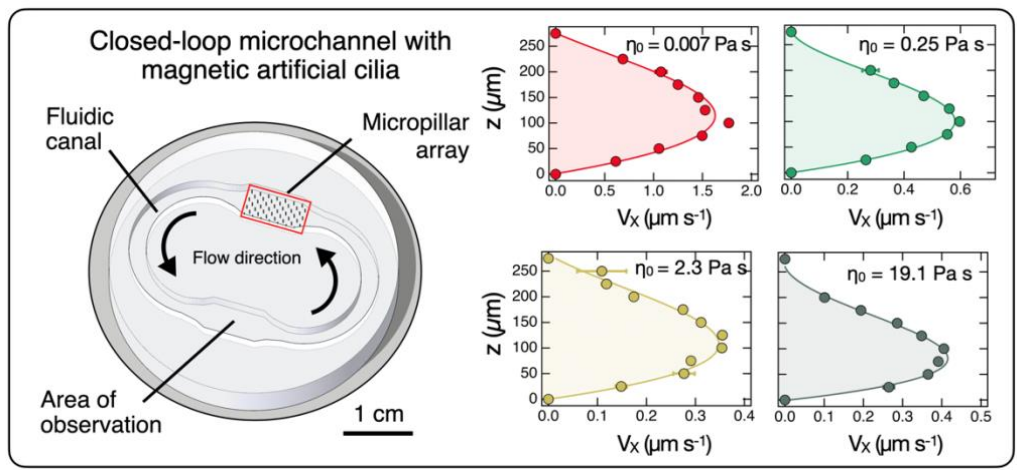